\documentclass[journal]{IEEEtran}
\usepackage[pdftex]{graphicx}
\usepackage{amsmath}
\usepackage{bm} 
\usepackage{mathrsfs}
\usepackage{amssymb}
\usepackage{clrscode}
\usepackage{setspace}
\usepackage{graphicx}
\usepackage{epstopdf}
\usepackage{epsfig}
\DeclareMathOperator*{\argmax}{arg\,max}
\begin{document}
%\begin{spacing}{2.0}
\title{Power vs. Spectrum 2-D Sensing in Energy Harvesting Cognitive Radio Networks }%\footnote{Part of the result has been submitted to ICC'15 Workshop on Green Communications and Networks with Energy Harvesting, Smart Grids, and Renewable Energies.}}

%\author{
%    \IEEEauthorblockN{Yanyan Zhang\IEEEauthorrefmark{1}, Weijia Han\IEEEauthorrefmark{2}, Di Li\IEEEauthorrefmark{1}, Ping Zhang\IEEEauthorrefmark{3} and Shuguang Cui\IEEEauthorrefmark{1}}
   
% \IEEEauthorblockA{\IEEEauthorrefmark{1}Texas A\&M University, \IEEEauthorrefmark{2}Xidian University, \IEEEauthorrefmark{3}Beijing University of Posts and Telecommunications}}

\author{Yanyan~Zhang,~\IEEEmembership{Student Member,~IEEE,}
        Weijia~Han,~\IEEEmembership{Member,~IEEE,}
        Di~Li,~\IEEEmembership{Student Member,~IEEE,}
        Ping~Zhang,~\IEEEmembership{Member,~IEEE,}
        and~Shuguang~Cui,~\IEEEmembership{Fellow,~IEEE}% <-this % stops a space

\thanks{The work was supported in part by DoD with grant HDTRA1-13-1-0029, by NSF with grants CNS-1343155, ECCS-1305979, and CNS-1265227, and by grant NSFC-61328102. }
\thanks{Y. Zhang, D. Li and S. Cui are with the Department of Electrical and Computer Engineering, Texas A\&M University, College Station, TX 77843, USA (e-mail:yanyan1016@email.tamu.edu; dili@tamu.edu; cui@tamu.edu). S. Cui is also a Distinguished Adjunct Professor at King Abdulaziz University, Jeddah 22254, Saudi Arabia.}
\thanks{W. Han is with the Information Science Institute, Xidian University, Xian 710126, China. (email:alfret@gmail.com)}
\thanks{P. Zhang is with the Wireless Technology Innovation Institute, Beijing University of Posts and Telecommunications, Beijing 100876, China. (email: pzhang@bupt.edu.cn) }
\thanks{Part of the result has been submitted to ICC'15 Workshop on Green Communications and Networks with Energy Harvesting, Smart Grids, and Renewable Energies.}% <-this % stops a space
}

\maketitle

\begin{abstract}
Energy harvester based cognitive radio is a promising solution to address the shortage of both spectrum and energy. Since the spectrum access and power consumption patterns are interdependent, and the power value harvested from certain environmental sources are spatially correlated, the new power dimension could provide additional information to enhance the spectrum sensing accuracy. In this paper, the Markovian behavior of the primary users is considered, based on which we adopt a hidden input Markov model to specify the primary vs. secondary dynamics in the system. Accordingly, we propose a 2-D spectrum and power (harvested) sensing scheme to improve the primary user detection performance, which is also capable of estimating the primary transmit power level. Theoretical and simulated results demonstrate the effectiveness of the proposed scheme, in term of the performance gain achieved by considering the new power dimension.  To the best of our knowledge, this is the first work to jointly consider the spectrum and power dimensions for the cognitive primary user detection problem.
\end{abstract}


\section{Introduction}
With the rapid development of wireless services in the past few
decades, spectrum resource, which is vital and hard-limited, faces a critical situation of being 
scarce. However, studies have revealed that we are wasting the spectrum as most allocated bands are being under-utilized. To address this problem, researchers have proposed
 the idea of dynamic spectrum  access, which could help
increase spectrum efficiency.

Cognitive Radio (CR) is the well-accepted technology to achieve dynamic spectrum
 access, with its core idea  of allowing Secondary Users (SUs) to access
spectrum when the licensed Primary Users (PUs) are idle.
The goal for CRs is to maximize the overall spectrum
efficiency while preventing harmful interference to PU transmissions.
One crucial building block of CR is spectrum sensing,
which determines whether certain spectrum is occupied by some active PUs.

Many statistical methods have been adopted for spectrum sensing in cognitive radio design. The authors in \cite{4453893} apply the Neyman-Perason lemma to study both the local and cooperative PU detection schemes, in which the sufficient statistics is compared against a certain threshold to detect the channel status. When the PU transmission signaling is known at the SU side, cyclostationary features could be explored for PU detection \cite{4387507}. For wideband cognitive radio systems, in addition to energy detection  \cite{4533213}, compressive sensing \cite{4218361} is also adopted to efficiently identify spectrum holes. However, most of methods above are sensitive to the shadowing/fading effects over the PU-to-SU link.

On the other hand, there are many related works that explore the PU Markovian behavior.   The existence of PU Markov patterns is validated in \cite{4912868} by real-time measurements in the paging band (928-948 MHz), with different effects of false alarm and miss detection are studied in \cite{5587873}, where a modified forward-backward detection algorithm is proposed to minimize the detection risks. In \cite{6575086}, a Hidden Bivariate Markov Model (HBMM) is used to quantify both the channel status and its dwelling time. For collaborative spectrum sensing, a parameter estimation algorithm with classification method is introduced in \cite{6635249} to identify the malicious users based on a Hidden Markov Model (HMM). 

Meanwhile, due to the growing demand of energy efficiency, the energy harvesting based cognitive radio network emerges to both improve channel utilization and meet the requirement of green communications. In the power domain,  Markov chain models have been widely accepted \cite{6183387}\cite{4586054} to specify the  
energy arrival process, for which some other statistical models have also been applied. For example, a Poisson model with a  known intensity $\lambda_0$ is adopted in \cite{6555257} and  a Gamma distribution  model is adopted in \cite{6449245}. Meanwhile, for wind energy harvesting systems, Weibull distribution \cite{6504247} is widely adopted  to forecast the wind speed and the corresponding harvested power level.

Unlike traditional communication systems, for the ones powered by the environment energy harvesters, power becomes a multiple access medium since the power usages across different users (especially geographically neighboring ones) are correlated \cite{6175764}\cite{CidFuentes}. Accordingly, the spectrum access and power access are actually correlated events,  and such correlations could be explored to help with improving the PU detection performance in CR, which used to be solely dependent on spectrum sensing.
In this work, we  consider the PU detection problem in energy harvesting based cognitive radio networks.  Given the correlation between spectrum and power usages, and by considering the spatial correlation among energy harvesting users, we propose a 2-D sensing scheme that could infer the PU behaviors by jointly learning the spectrum and power access dynamics. This is promising to improve the PU detection performance since the traditional methods, which are solely based on spectrum sensing,  have certain limitation: When the PU-to-SU transmission is under fading/shadowing, the detection performance degrades sharply as the channel observation is no longer reliable. Since the PU-to-SU channel quality does not affect the power-dimension inference, the proposed 2-D scheme could overcome the effect of channel fading/shadowing  and provide a more reliable combined sensing result. 

In addition, traditional spectrum sensing methods  only focus on sensing the ``on-off" activity of PUs.  There is also a growing demand of knowing the transmission power levels of PUs. For example, in \cite{5054703},  a novel method that utilizes not only the temporal but also the spatial spectrum holes is proposed, which requires the knowledge of PU transmission power to estimate the PU coverage area. In \cite{6477928}, an optimal SU power allocation scheme  is provided, which also depends on the knowledge of multiple PU transmission power levels. As a side product, the 2-D sensing scheme proposed in this paper could sense the spectrum and estimate the PU transmit power level simultaneously, which could provide more potentials to enhance the performance of the energy harvesting based CR networks. 

The rest of this paper is organized as follows. Section \uppercase\expandafter{\romannumeral 2} describes the basic system model. In Section \uppercase\expandafter{\romannumeral 3}, the Hidden Input Markov Model (HIMM) is proposed to further specify the system dynamics.  In Section \uppercase\expandafter{\romannumeral 4}, the proposed 2-D sensing scheme is analyzed. In Section \uppercase\expandafter{\romannumeral 5}, the parameter learning algorithm for HIMM is studied. In Section \uppercase\expandafter{\romannumeral 6}, the simulation results are provided to show the advantage of the proposed scheme. Finally, Section \uppercase\expandafter{\romannumeral 7} concludes the paper.

\section{System Model}
We consider a simple cognitive system with one PU and one SU, which are both powered by harvested energy from the same renewable source. If the PU does not transmit at  certain time slots, the harvested energy is discarded since no battery is assumed \cite{6190022}.  In our setup, it is possible that even when the PU has data, the energy level may not meet the  reliable transmission minimum requirement, which implies that the PU does not occupy the channel. This is the extra PU idle case compared with the traditional PU system where PU is idle only if it has no data to transmit. Specifically, let $H_0$ denote the case when the PU does not transmit and $H_1$ denote the case when the PU occupies the channel; then a formal definition for $H_0$ and $H_1$ in an energy harvesting based cognitive radio network is:
 $$H_0: No\ Enough\ Energy \text{ or } No\ Data\ Available\ ;$$
$$H_1: Enough\ Energy \text{ and } Data\ Available.$$

Let $\{ E_t, t=1,2,3,\cdots \}$ denote the energy arrival process at the PU on a time-slotted basis, where each $E_t$ could take value from a set with a finite cardinality  $\mathscr{L}=\{i, i+1, i+2 , \cdots, i+L-1\}$, $ i \geq 0$, i.e., the harvested energy value is quantized into $L$ levels. Here $L$ is set as a relatively large number, such that the effect of discretizing the PU energy level is negligible. 
The channel occupancy state $\{C_t, t=1, 2,3, \cdots\}$ is a discrete-time process such that  each state takes value in $\mathscr{C}=\{0,1,2,\cdots,M-1\}$, with $M=2$ in this paper, which implies: When $C_t=0$ the channel is idle, and when $C_t=1$ the channel is occupied by the PU.

 Let $E_{\text{h}}$ be the minimum energy level required for reliable primary transmission and $P_0$ be the probability of no data available. Then the relationship between the channel state $\{ C_t\}$ and the PU energy level $\{ E_t \}$ could be expressed as 
\begin{flalign}\label{scheme}
P(C_t=0)&=P(E_t< E_{\text{h}})+P(E_t \geq E_{\text{h}})\cdot P_0, \nonumber \\
P(C_t=1)&=P(E_t \geq E_{\text{h}})\cdot (1-P_0).
\end{flalign}

Note that we assume no battery installed such that when a PU transmits, it uses up all the harvested energy available at that time slot; when a PU does not transmit, it discards the harvested energy in that time slot. 
On the SU side, since the channel state and PU energy level are not directly observable, SU could only estimate the hidden states with its available observations. First, as both PU and SU are powered by harvested energy from the same renewable source, the harvested SU energy could be treated as an observation for the PU energy level, due to the spatial correlation of the energy harvesting processes.  The relationship between the latent PU harvested energy $E_t$ and the SU harvested energy $U_t$ will be further discussed in Section \uppercase\expandafter{\romannumeral 3} (See Fig. \ref{Fig1}).

Assuming that the PU and SU operate on a synchronous time-slotted fashion, the sampled received signal from PU to SU at time slot $t$ is given as (under real-valued signaling assumption)
\begin{flalign}\label{signal}
\begin{cases}
\text{H}_0: \quad x_t(n)=u_t(n), \quad n=1,2, \cdots, N,\\
\text{H}_1: \quad x_t(n)=h_t\cdot s_t(n)+u_t(n),\quad n=1,2, \cdots, N,
\end{cases}
\end{flalign}
where $n$ is the sampling index with a total of $N$ samples per slot, $s_t(n)$ is the signal transmitted by the PU, $h_t$ is the channel gain constant over each slot, and $u_t(n)$ denotes the i.i.d. Gaussian noise. The signal sample $s_t(n)$ is assumed i.i.d and independent from the noise. Denote $Y_t$ as the summation of $N$ received signal energy samples at the SU, i.e., 
\begin{flalign}\label{samples}
Y_t= \sum_{i=0}^{N-1}x_t^2(i). 
\end{flalign}

In the following sections, we will first model the mathematical structure of the SU observation signals and then learn the parameters of HIMM, based on which the  
 PU hidden states  are estimated. 

\section{Hidden Input Markov Model}
\begin{figure}[!t]
\centering
\epsfxsize=3.5in
\epsfbox{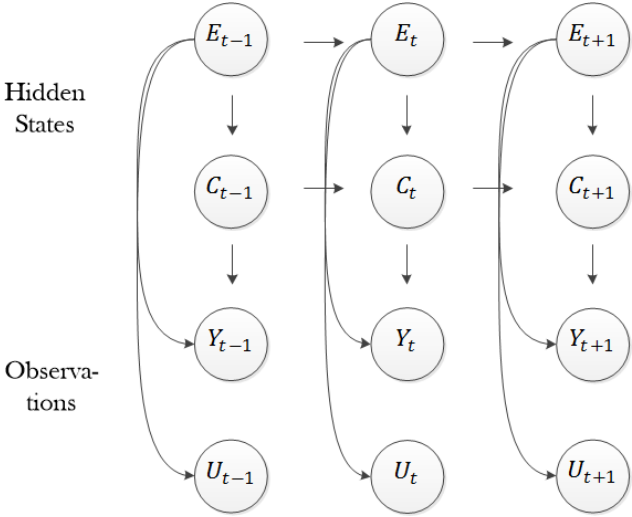}
\caption{Hidden Input Markov Model Structure}
\label{Fig1}
\end{figure}

Markov chain has been widely adopted \cite{6183387}\cite{4586054} to specify the environmental energy harvesting process, and to model the PU data arrival process in traditional cognitive radio networks \cite{4912868}\cite{6575086}.
In our work here, we first adopt two discrete-time Markov chains to represent the PU data arrival and energy harvesting processes respectively, then quantify the 2-D signaling structure with a HIMM as shown in Fig. \ref{Fig1}.

From Fig. \ref{Fig1}, we see that there are several differences between the traditional HMMs and the Markov structure abstracted in our problem. If the PU energy level is known, which implies that $U_t$ reflects $E_t$ perfectly, then the structure of our problem becomes similar to the Input-Output Hidden Markov Model (IOHMM) \cite{536317}.  For IOHMM, the training process  is a supervised learning problem, such that with the training input and output, the functional mapping between the input and the output could be inferred. Our structure is more complicated than IOHMM, since the unobservable Markov state $ C_t$ is a function of both the previous state $C_{t-1}$ and the input $ E_t$, while the input $E_t$ is also a hidden Markov process from the SU's point of view. On the other hand, the hidden input could be observed not only from the SU energy level $U_t$, but also from the channel observation $Y_t$ when the channel is active. We thus call this structure as HIMM, for which the existing HMM results could not be directly applied.

As  both $\{E_t\}$ and $\{ C_t\}$ are first-order Markov chains,  according to \eqref{scheme} we have $C_t=f(C_{t-1},E_t)$. Let $C^t$ denote a collection of channel states from time $1$ up to time $t$ (with a similar definition for $E^t$). The first-order Markov property implies the following relationship
\begin{flalign}
P(C_t|C^{t-1},E_t)&=P(C_t|C_{t-1},E_t), \nonumber \\
P(E_t|E^{t-1})&=P(E_t|E_{t-1}).
\end{flalign}
Throughout the paper the notation $P(A)$ always refers to the probability of the random variable taking a particular value, i.e., $P(A)=P(A=a)$, unless specified otherwise.

We use $\bold{A}$ and $\bold{B}$ to denote the transition matrices for the processes of $\{ E_t\}$ and $\{ C_t\}$, respectively. Clearly, as $\{E_t\}$ is a traditional Markov chain and takes values from $L$ states, $\bold{A}=\{a_{ij}\}$ is a $L \times L$ matrix, in which each component $a_{ij}$ is given as
\begin{flalign}
a_{ij}=P(E_t=j|E_{t-1}=i), \quad i,j \in \mathscr{L}.
\end{flalign} 

The transition matrix of $\bold{B}=\{\bold{B}(q), q \in \mathscr{L}\}$ is actually a set of  matrices. For each $q$, $\bold{B}(q)$ is a $M \times M$ matrix. Inside $\mathbf{B}(q)$, each component $b_{ij,q}$ indicates the following transition relationship of channel states:
\begin{flalign}
b_{ij,q}=P(C_t=j|C_{t-1}=i,E_t=q), \quad i,j \in \mathscr{C},  q \in \mathscr{L}.
\end{flalign}

Besides the transition probability, the initial probability distribution is also important in describing a Markov chain. Here we use vector ${\pi}^E$ and matrix $\boldsymbol{\pi}^C$ to specify the  initial distributions, in which each element stands for 
\begin{flalign}
{\pi}_{i}^E&=P(E_1=i),  \quad i \in \mathscr{L} \nonumber \\
\pi_{ij}^C&=P(C_1=j|E_1=i), \quad i \in \mathscr{L}, j \in \mathscr{C}.
\end{flalign}

For the HIMM shown in Fig. \ref{Fig1}, the SU observations include the SU energy level $U_t$ and the SU received signal energy $Y_t$. 
 Mathematically, due to spatial correlation, we could model $U_t$ as a function of $E_t$. Without loss of generality, we assume $U_t \in \mathscr{L}$.  Let the $L \times L$ matrix $\bold{D}= \{d_{ij} \}$ be the emission matrix with each $d_{ij}$ defined as
\begin{flalign}
d_{ij}=P(U_t=j|E_t=i), \quad i,j\in \mathscr{L}.
\end{flalign}
Note that conditioned on $E_t$, the only randomness left at $U_t$ is the measurement noise, which we could assume independent over time. Therefore, we could have the following decomposition:
\begin{flalign}
P(U^t|E^t)=\prod_{i=1}^{t}P(U_i|E_i).
\end{flalign}

On the other hand, based on the Central Limit Theorem (CLT), when the number of received signal samples $N$ is relatively large, the sum  in \eqref{samples} follows a Gaussian distribution. Therefore, conditioned on the hidden states, we assume that the observation $\{ Y_t\}$ follows a Gaussian  distribution, having its relationship with $E_t$ and $C_t$ shown in Fig. \ref{Fig1}. According to the SU received signal in \eqref{signal}, when the channel state is idle, the output $Y_t$ only reflects the energy of channel noise, which contains no information about the PU energy level; when the channel is busy, the channel output includes the PU signal, such that the distribution of $Y_t$ is affected by both the hidden states $E_t$ and $C_t$. For each $t$, $Y_t$ is  distributed as 
\begin{flalign}
P(Y_t|C_t=i,E_t=j) \sim N(\mu_{ij}, \sigma^2_{ij}), \quad i \in \mathscr{C}, j\in \mathscr{L}.
\end{flalign}

Based on our previous discussion, when $i = 0$, $\mu_{ij}$ and $\sigma^2_{ij}$ are respectively identical for all $j$'s. Let matrix $\boldsymbol{\mu}=\{\mu_{ij}\}$ and matrix $\boldsymbol{\sigma}^2=\{ \sigma^2_{ij}\}$, the emission probability of $Y_t$ is then specified by $\boldsymbol{\mu}$ and $\boldsymbol{\sigma}^2$. Also, given the hidden states, the channel observation $Y_t$ is conditionally independent over time.

The initial probability, transition probability, and emission probability are three important parts that constitute the parameters of this unobservable Markov chain. Let $\eta=\{ {\pi}^E, \boldsymbol{\pi}^C, \bold{A}, \bold{B}, \bold{D}, \boldsymbol{\mu}, \boldsymbol{\sigma}^2 \}$ be the collection of these parameters. When $\eta$ is decided, the entire structure of this Markov model is specified. Therefore, learning these parameters is crucial in order for us to explore the application of this unobservable Markov chain, which is used in our PU detection scheme.

\begin{figure}[!t]
\centering
\epsfxsize=3.5in
\epsfbox{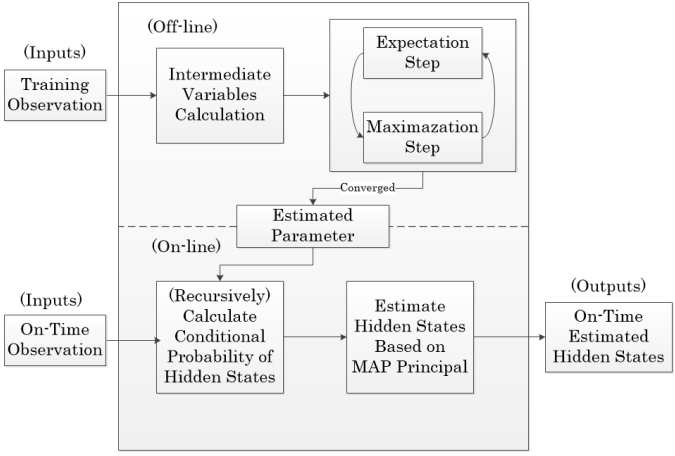} 
\caption{Structure of the Proposed Algorithm}
\label{sequence}
\end{figure}

The structure of the overall scheme is provided in Fig.  \ref{sequence} to better illustrate the work. In the following two sections, we first discuss the PU detection scheme assuming an estimated parameter $\eta'$; then we present a parameter learning algorithm. The arrangement will make the flow more smooth since the parameter learning algorithm is quite lengthy.

\section{2-D Sensing}
In this section, we analyze the PU detection scheme with 2-D sensing over the HIMM model. Let $t$ denote the current time slot; since HIMM takes the past observations into consideration, at time $t$ the past and current observations $\{ U^t, Y^t\}$ are all available to contribute to the detection process. With the estimated parameter vector $\eta'$, the probability measure of the current PU hidden states could be obtained by recursively marginalizing all the past states, while the observations in each time slot work as correctors to modify the previously predicted result.

\subsection{Sensing Algorithm}
First, we state the 2-D sensing scheme as follows. The detection for the hidden PU channel state $\hat{C}_t$ and the estimation for the PU energy level $\hat{E}_t$ could be jointly achieved by maximizing the conditional probability of the current hidden states, i.e., $(\hat{C}_t, \hat{E}_t)$ is decided as
 \begin{flalign}\label{2D}
\hat{C}_t, \hat{E}_t= \argmax_{c_t,e_t} \quad P(C_t=c_t, E_t=e_t| U^t, Y^t).
\end{flalign}
In CR, such an estimation result further decides the activity of SU: When $\hat{C}_t=0$, SU considers the channel as inactive and utilizes it; when $\hat{C}_t=1$, SU considers the channel as occupied and stays silent or transmits carefully based on the knowledge of $\hat{E}_t$. 

In fact, the joint probability  $P(U^t, Y^t)$ is the proportional constant between $P(C_t, E_t| U^t, Y^t)$ and $P(C_t, E_t,U^t, Y^t)$, i.e., by Bayes rules there is a relationship
\begin{flalign}\label{prop}
P(C_t, E_t| U^t, Y^t)=P(C_t, E_t,U^t, Y^t)/P(U^t, Y^t) \nonumber \\
\varpropto P(C_t, E_t,U^t, Y^t).
\end{flalign}
This result implies that the estimated hidden states $\hat{C}_t$, $\hat{E}_t$ also maximize the joint probability $P(C_t, E_t,U^t, Y^t)$. In \cite{Kay:1993:FSS:151045}, it is demonstrated that such estimated results are based on the Maximum A Posteriori (MAP) principle, which is considered as an optimal estimate to minimize the Bayes risk for a ``hit-or-miss" cost function. Next we discuss how to evaluate and solve \eqref{2D}.

We first give the procedure to recursively calculate $ P(C_t, E_t| U^t, Y^t)$  in \eqref{2D}, which is divided into the following three steps.  
Since the Markovian property could not be easily applied to decompose the conditional probability of the hidden states for recursive calculations, in the first step, the probability of the current observation given the past $P(U_t,Y_t|U^{t-1},Y^{t-1})$ is used to multiply the target conditional probability  $P(C_t, E_t| U^t, Y^t) $, such that the production after multiplication could be directly decomposed and computed as 
{\small\begin{flalign}\label{divide_1}
&\quad P(U_t,Y_t|U^{t-1},Y^{t-1})P(C_t, E_t| U^t, Y^t) \nonumber \\
&=\sum_{C_{t-1}}\sum_{E_{t-1}} P(C_t, C_{t-1}, E_t, E_{t-1},U_t, Y_t| U^{t-1}, Y^{t-1}) \nonumber \\
&\overset{\text{(a)}}{=}\sum_{C_{t-1}}\sum_{E_{t-1}} P(Y_t|C_t, E_t)P(U_t|E_t) \nonumber \\
&\qquad \qquad \qquad \bm\cdot P(C_t, C_{t-1},E_t, E_{t-1} |U^{t-1}, Y^{t-1}) \nonumber \\
&=\underbrace{P(Y_t|C_t, E_t)P(U_t|E_t)}_{\text{corrector}}\bm\cdot  \nonumber \\
&  \underbrace{ \sum_{C_{t-1}}\sum_{E_{t-1}}P(C_t|C_{t-1},E_t)P(E_t|E_{t-1})P(C_{t-1},E_{t-1}|U^{t-1},Y^{t-1})}_{\text{predictor}},
\end{flalign}}
where (a) is obtained with the Markovian property and Bayes rules, and the current observations $Y_t$ and $U_t$ work as  correctors to improve the prediction of the current states based on the past. 
Inside \eqref{divide_1}, the probabilities of the corrector part could be computed by the emission matrices from the parameter vector $\eta'$, the first two probabilities of the predictor could be computed by the transition matrices in $\eta'$, and the last probability is the previous conditional probability of the hidden states, which is obtained from the previous step in our recursive algorithm. 
Note that the recursion procedure is exactly connected by the last probability term inside \eqref{divide_1}, as it is the only term dependent on the previous information. 

After we compute \eqref{divide_1}, to obtain the objective in \eqref{2D}, the value of $P(U_t,Y_t|U^{t-1},Y^{t-1})$ has to be calculated in order to divide it out from \eqref{divide_1}. 
Thus the second step calculates $P(U_t,Y_t|U^{t-1},Y^{t-1})$: As the summation of $P(C_t, E_t| U^t, Y^t)$ over all possible hidden states equals one, the value of $P(U_t,Y_t|U^{t-1},Y^{t-1})$ could be calculated as
{\small\begin{flalign}\label{divide_2}
&\quad P(U_t,Y_t|U^{t-1},Y^{t-1})  \nonumber \\
&=P(U_t,Y_t|U^{t-1},Y^{t-1})\sum_{C_{t}}\sum_{E_{t}}P(C_t, E_t| U^t, Y^t) \nonumber \\
&=\sum_{C_{t}}\sum_{E_{t}}P(U_t,Y_t|U^{t-1},Y^{t-1})P(C_t, E_t| U^t, Y^t).
\end{flalign}}
Although the individual probability terms inside \eqref{divide_2} could not be directly computed, the multiplication of the  two probabilities actually is already the result we obtained in \eqref{divide_1} in our first step. Then marginalizing over all possible $(C_t,E_t)$ pairs leads to the conditional probability of current observations $P(U_t,Y_t|U^{t-1},Y^{t-1})$, as we desire.

At the last step, the conditional probability of hidden states $P(C_t, E_t| U^t, Y^t)$ is calculated by dividing \eqref{divide_1} with \eqref{divide_2}, which actually could be treated as a normalization procedure. It is worth noting that when $t=1$, which is the starting point of our recursion algorithm, the multiplication corresponding to \eqref{divide_1} is expressed as
\begin{flalign}\label{initial}
& \quad P(U_1,Y_1)P(E_1,C_1|U_1,Y_1) \nonumber \\
&=P(E_1,C_1,U_1,Y_1) \nonumber \\
&=P(U_1,Y_1|E_1,C_1)P(E_1,C_1) \nonumber \\
&=P(U_1|E_1)P(Y_1|E_1,C_1)P(C_1|E_1)P(E_1),
\end{flalign} 
where the first two probabilities are from the emission matrices in the parameter vector $\eta'$ and the last two probabilities are from the initial probabilities $\boldsymbol{\pi}^C$ and ${\pi}^E$ inside $\eta'$. Using the same method as in $\eqref{divide_2}$ to obtain the value of $P(U_1,Y_1)$ and dividing the multiplication result in \eqref{initial} by $P(U_1,Y_1)$, the conditional probability of the hidden states at $t=1$ could be computed.

After obtaining the conditional probability of the hidden states $P(C_t, E_t| U^t, Y^t)$, the maximization procedure in \eqref{2D} could be solved by a simple 2-D exhaustive search, since the cardinality of the possible hidden states are within a finite range.

In summary, the algorithm to compute the conditional probability $P(C_t, E_t| U^t, Y^t)$ is summarized below.
\begin{codebox}
\Procname{$\proc{Algorithm to Compute $P(C_t, E_t| U^t, Y^t)$ }$}
 \textbf{Input:} $\eta$, $U^t$,  $Y^t$ \\
\li    for $i\in \mathscr{L}$
\li      \quad for $j \in \mathscr{C}$
\li    \qquad $P(E_1=i,C_1=j|U_1,Y_1)= \frac{Q(1)\pi^c_{ij}\pi^E_{i}}{\sum\limits_{i \in \mathscr{L}}\sum\limits_{j \in \mathscr{C}} Q(1)\pi^c_{ij}\pi^E_{i}}$
\li      \quad end
\li       end
\li for $k=2,\cdots,t$
\li \quad for $i \in \mathscr{L}$
\li \qquad for $j \in \mathscr{C}$
\li    \qquad \quad $P(E_k=i,C_k=j|U^k,Y^k)=$ \\
   ${\frac{Q(k)\sum\limits_{l \in \mathscr{L}}\sum\limits_{m \in \mathscr{C}} b_{mj,i}a_{li}P(C_{k-1}=m,E_{k-1}=l|U^{k-1},Y^{k-1})     }{\sum\limits_{i \in \mathscr{L}}\sum\limits_{j \in \mathscr{C}} Q(k) \sum\limits_{l \in \mathscr{L}}\sum\limits_{m \in \mathscr{C}} b_{mj,i}a_{li}P(C_{k-1}=m,E_{k-1}=l|U^{k-1},Y^{k-1})   }}$
\li   \qquad end
\li      \quad end
\li       end
\end{codebox}
where $Q(k)$ stands for $d_{i U_k}\frac{\exp(-\frac{(Y_k-\mu_{ji})^2}{2\sigma_{ji}^2})}{\sqrt{2 \pi \sigma^2_{ji}}}$.

\emph{Remark:} For an energy harvesting based PU, there is always a collection of energy states that are not enough to support reliable PU transmissions. Let $\mathscr{L}_0 \subset \mathscr{L}$ denote the subset that contains all the energy states from $\mathscr{L}$ that are insufficient for reliable PU transmissions and $\mathscr{L}_1$ be the complement of $\mathscr{L}_0$. From the relationship given in \eqref{scheme}, when $\hat{E}_t \in \mathscr{L}_1$, the estimated channel state $\hat{C}_t$ could be either $1$ or $0$. However, when $\hat{E}_t \in \mathscr{L}_0$,  $\hat{C}_t$ could only be zero. This is due to the fact that the joint probability of $C_t=1$ and $E_t \in \mathscr{L}_0$ is always equal to zero.  As such, the optimal solution for \eqref{2D} could not be $(\hat{C}_t=1,\hat{E}_t \in \mathscr{L}_0)$, which could be eliminated from the searching space in solving $\eqref{2D}$ to reduce the computational complexity.

\subsection{Performance Comparison via Mutual Information}
Compared with our proposed 2-D sensing scheme, the traditional PU detection methods do not take the spatial correlation of energy harvesting processes into consideration, i.e., for traditional methods the only available observation is $Y^t$. It implies that under the same MAP principle, the traditional sensing scheme is basically solving 
\begin{flalign}\label{1D}
\hat{C}_t, \hat{E}_t=\argmax_{c_t,e_t}P(C_t=c_t,E_t=e_t,Y^t).
\end{flalign}

The mutual information between the hidden states and observations could be used to quantify the estimation performance. In our setup, it can be evaluated as 
\begin{flalign}
I(C_t,E_t;Y^t,U^t)=I(C_t,E_t;Y^t)+I(C_t,E_t;U^t|Y^t),
\end{flalign}
where $I(C_t,E_t;Y^t)$ is the mutual information quantitying the performance of the traditional detection methods, and $I(C_t,E_t;U^t|Y^t)$ is the information gained by additionally considering the power domain information. The mutual information gain is zero if and only if $(C_t,E_t)$ are independent of $U^t$ conditioned on $Y^t$, which is not the case in our system, such that the mutual information gain is always greater than zero in our method.

The above mutual information gain demonstrates that $U^t$ could help the estimation of the PU hidden states; therefore, the proposed 2-D sensing method has the potential to perform better than the traditional detection methods, which will be further validated by simulation results in Section \uppercase\expandafter{\romannumeral 6}. The benefit of using energy observations for PU detection could be much more remarkable when the PU-to-SU links experience deep fading or shadowing, where the channel observations are much less reliable but energy observations are not affected by such channel impairment. Simulation results in Section \uppercase\expandafter{\romannumeral 6} will further illustrate this point.

\section{Parameter Estimation}
In this section, we discuss the proposed algorithm for parameter estimation. We need to first define several intermediate variables, which are used for algorithm implementation, then derive the algorithm based on the traditional EM algorithm. Note that the proposed algorithm is off-line, with observations $U^T$, $Y^T$ for training.

\subsection{Intermediate Variables}
 For $U^t$ and $E_t$, define 
\begin{flalign}
\alpha_i^U(t) \triangleq P(U^t, E_t=i|\eta), \quad t=1, 2, 3, \cdots, T,
\end{flalign} 
which is the joint probability of getting the observation $U^t$ and having the PU energy state at time $t$ ($E_t$) equal $i$, conditioned on $\eta$, with $\eta$ defined in Section \uppercase\expandafter{\romannumeral 3}. Similarly, with the communication channel output observation, define 
\begin{flalign}
\alpha_{ij}^Y(t) \triangleq P(Y^t, C_t=i, E_t=j|\eta), \quad t=1, 2, 3, \cdots, T,
\end{flalign}  
which is the joint probability of getting the observations $Y^t$ and having the channel state $C_t$ as $i$ and the energy level $E_t$ as $j$, conditioned on $\eta$ as well.

Since we estimate two hidden states with different observations, the information carried by the joint observations should also be considered. Therefore, define
\begin{flalign}
\alpha_{ij}^{U,Y}(t) \triangleq P(U^t,Y^t, C_t=i, E_t=j|\eta), \quad t=1, 2, 3, \cdots, T,
\end{flalign} 
as the intermediate variables for joint observations, which indicates the probability of getting the observations $U^t$, $Y^t$, with $C_t=i$ and $E_t=j$.

As both $\{ E_t \}$ and $\{ C_t\}$ are Markov processes, with Markov property, the defined intermediate variable collections $\alpha^U$, $\alpha^Y$ and $\alpha^{U,Y}$ can be quantified recursively. For $\alpha^U$, the probability at time $t+1$ is given as 
\begin{flalign}
%\alpha_i^E(1)& = \pi_{Ei}f(U_1|i), \nonumber \\
\alpha_j^U(t+1) = f(U_{t+1}|j)\sum_{i \in \mathscr{L}}\alpha_i^U(t)a_{ij} \quad t=2,\cdots,T,
\end{flalign} 
which is traced to $\alpha_i^U(1)= \pi_{Ei}d_{iU_1}$ back at the initial state.
This process is usually called a forward recursion, in which the next-time joint probability equals to the  current joint probability weighted by the transition probability, followed by marginalization and multiplication with the emission probability. It can be interpreted that the forward recursion at each time equals to a predictor multiplied by a corrector. 
 Similarly, the forward procedures for $\alpha^Y$ and $\alpha^{U,Y}$ are respectively given as
\begin{flalign}
\alpha_{ij}^Y(1)& = \pi_{Cji}\pi_{Ej}f(Y_1|i,j), \nonumber \\
\alpha_{ij}^Y(t+1)& = f(Y_{t+1}|i,j)\sum_{l \in \mathscr{L}} \sum_{m \in \mathscr{Y}}\alpha_{ml}^Y(t)a_{lj}b_{mi,j}, \nonumber \\
&\qquad \quad \qquad\qquad\qquad t=2,\cdots,T.
\end{flalign} 
and 
\begin{flalign}
\alpha_{ij}^{U,Y}(1)& =\pi_{Cji}\pi_{Ej}d_{jU_1}f(Y_1|i,j), \nonumber \\
\alpha_{ij}^{U,Y}(t+1)& = d_{jU_1}f(Y_{t+1}|i,j)\sum_{l \in \mathscr{L}} \sum_{m \in \mathscr{C}}\alpha_{ml}^{U,Y}(t)a_{lj}b_{mi,j}, \nonumber \\
&\qquad \quad \qquad\qquad\qquad t=2,\cdots,T.
\end{flalign} 

In addition to the forward recursion, backward recursions are then used to measure the probability of getting future output up to time $T$ given the state at time $t$. Define
\begin{flalign}
\beta_i^U(t) &\triangleq P(U_{t+1},U_{t+2},\cdots,U_T|E_t=i,\eta), \nonumber \\
\beta_{ij}^Y(t) &\triangleq P(Y_{t+1},Y_{t+2},\cdots,Y_{T}|C_t=i,E_t=j,\eta), \nonumber \\
\beta_{ij}^{U,Y}(t) &\triangleq \nonumber \\
 P(U_{t+1}&,Y_{t+1},U_{t+2},Y_{t+2},\cdots,U_{T},Y_{T}|C_t=i,E_t=j,\eta), \nonumber \\
&\qquad \qquad \qquad \qquad t= 1, 2, 3, \cdots, T,
\end{flalign} 
as the probabilities of getting the observation sequence from $t+1$ to $T$, conditioned on the hidden states at time $t$. For the SU energy state, $\beta_i^U(t)$ is calculated as
\begin{flalign}
\beta_i^U(t)= \sum_{j\in \mathscr{L} }\beta_j^U(t+1)a_{ij}d_{jU_{t+1}}, \quad t=1,\cdots,T-1, 
\end{flalign}
with $\beta_i^U(T)=1, \ \forall i$. This indicates that the probability of getting the future observations up to time $T$ conditioned on the current hidden states equals to such a conditional probability at the next time step multiplying the transition and emission probabilities of all possible current latent states. As it can be seen, the backward recursion uses the future observations to smooth the current inference. 
Meanwhile, the calculations for $\beta_{ij}^Y(t)$ and $\beta_{ij}^{U,Y}(t)$ could be expressed as 
\begin{flalign}
\beta_{ij}^Y(t)&=\sum_{m\in \mathscr{C}}\sum_{l\in \mathscr{L}}\beta_{ml}^Y(t+1)f(Y_{t+1}|m,l)a_{jl}b_{im,l},\nonumber \\
\beta_{ij}^{U,Y}(t)&=\sum_{m\in \mathscr{C}}\sum_{l\in \mathscr{L}}\beta_{ml}^{U,Y}(t+1)f(Y_{t+1}|m,l)d_{lU_{t+1}}a_{jl}b_{im,l},\nonumber \\
&\qquad \quad \qquad\qquad\qquad \qquad t=1,\cdots,T-1,
\end{flalign}
with $\beta_{ij}^Y(T)=1, \beta_{ij}^{U,Y}(T)=1, \ \forall i,j$.

After obtaining the expressions for forward and backward recursions, we need to define two update variables for each observation process to make the implementation of the proposed parameter learning algorithm more straightforward. One update variable is defined to consider the joint probability of getting the entire observations from time $1$ to time $T$ and the hidden state at time $t$; the other one considers the joint probability of getting the entire observations and the hidden states at time $t$ and $t+1$. Specfically, for the SU energy observation process, define 
\begin{flalign}
\gamma_i^U(t) &\triangleq P(U^T, E_t=i|\eta)=\alpha^U_i(t)\beta_i^U(t), \quad t=1,\cdots,T \nonumber \\
\varepsilon_{ij}^U(t) &\triangleq P(U^T, E_t=i, E_{t+1}=j|\eta) \nonumber \\
& =\alpha^U_i(t) a_{ij}\beta_j^U(t+1)d_{jU_{t+1}}, \quad t=1,\cdots, T-1.
\end{flalign}
Clearly, the update variables above combine the forward and backward variables with transition and emission probabilities. For the channel observation and the joint observation processs, variables are defined as 
\begin{flalign}
\gamma_{ij}^Y(t) &\triangleq P(Y^T, C_t=i, E_t=j|\eta)=\alpha_{ij}^Y(t)\beta_{ij}^Y(t), \nonumber \\
\varepsilon_{ij,ml}^Y(t)& \triangleq P(Y^T, C_t=i, E_t=j, C_{t+1}=m, E_{t+1}=l|\eta) \nonumber \\
&=\alpha_{ij}^Y(t)a_{jl}b_{im,l}\beta_{ml}^Y(t+1)f(Y_{t+1}|m,l).
\end{flalign}
and
\begin{flalign}
\gamma_{ij}^{U,Y}(t) &\triangleq P(U^T,Y^T, C_t=i, E_t=j|\eta)=\alpha_{ij}^{U,Y}(t)\beta_{ij}^{U,Y}(t), \nonumber \\
\varepsilon_{ij,ml}^{U,Y}(t)& \triangleq P(U^T,Y^T, C_t=i, E_t=j, C_{t+1}=m, E_{t+1}=l|\eta) \nonumber \\
&=\alpha_{ij}^{U,Y}(t)a_{jl}b_{im,l}\beta_{ml}^{U,Y}(t+1)f(Y_{t+1}|m,l)d_{lU_{t+1}}.
\end{flalign}
Again, the reason to define these intermediate variables is that they will be used to implement the proposed parameter estimation algorithm for HIMM, which is introduced in the following subsection.  

\subsection{Parameter Estimation Algorithm}
The EM algorithm \cite{Bilmes98agentle} is adopted here used to iteratively find the unknown parameters that maximize the likelihood in a statistical model, in which the latent variables exist. Here, E stands for expectation and M stands for maximization, and the algorithm alternates between the E and M procedures. The current expectation step creates an expected log-likelihood function, which is averaged over the latent variables given the estimated parameters in the previous step; then the maximization step is designed to find a new parameter that maximizes the function of log-likelihood formulated in the current expectation step. The algorithm continues until the results converge (when the difference between the results of two  consecutive iterations is within an acceptable region). As proved before, the expected log-likelihood is non-decreasing during the iterations and the algorithm converges; however, it is not guaranteed to converge to a global optimum. This is due to the fact that although the maximization step is seeking the global optimum for the expected joint probability of hidden states and observations, the results may not be the global optimum for the likelihood function, which is from the conditional probability of observations. 
Detailed analysis about the EM algorithm convergence could be found in \cite{wu1983}. 

\subsubsection{Expectation Step}
In the expectation step, we need to consider the average of the log-likelihood function over the latent Markov states. If, in an extreme case, the Markov states are known, then the E step could be decomposed into $T$ number of subproblems since the temporal dependence of the Markov chain contains no additional information for the expectation of the log-likelihood function. Let $\eta^{(k-1)}$ denote the parameter estimated at the previous iteration; and let $O=\{U^T, Y^T\}$ stand for the collection of observations. The expectation of log-likelihood can then be expressed as
\begin{flalign}
&L(\eta;\eta^{(k-1)})= \mathbb{E}_{C^T,E^T} \{ \log P(E^T,C^T, O|\eta )\} \nonumber \\
&=\sum_{E^T \in \mathscr{L}^T}\sum_{C^T \in \mathscr{C}^T} P(E^T, C^T|O, \eta^{(k-1)})\log P(E^T, C^T, O|\eta) \nonumber \\
&=\sum_{E^T \in \mathscr{L}^T}\sum_{C^T \in \mathscr{C}^T} \frac{P(E^T, C^T, O|\eta^{(k-1)})}{P(O|\eta^{(k-1)})}\log P(E^T, C^T, O|\eta).
\end{flalign}
Inside the equation, $\sum\limits_{E^T \in \mathscr{L}^T}$ and $\sum\limits_{C^T \in \mathscr{C}^T}$  are abbreviations for $\sum\limits_{E_1\in \mathscr{L}}\sum\limits_{E_2\in \mathscr{L}}\cdots\sum\limits_{E_T\in \mathscr{L}}$ and $\sum\limits_{C_1\in \mathscr{C}}\sum\limits_{C_2\in \mathscr{C}}\cdots\sum\limits_{C_T\in \mathscr{C}}$. The achieved expected log-likelihood is utilized  in the Maximization step to find a better parameter estimate, which should not decrease the averaged log-likelihood.

\subsubsection{Maximization Step}%  {\pi}^E, \boldsymbol{\pi}^C, \bold{A}, \bold{B}, \bold{D}, \boldsymbol{\mu}, \boldsymbol{\sigma}^2
The objective in the M step is to achieve a better parameter estimate $\eta^{(k)}$, which is consisted of $\{\pi^{E(k)}, \boldsymbol{\pi}^{C(k)}, \bold{A}^{(k)}, \bold{B}^{(k)},  \bold{D}^{(k)}, \boldsymbol{\mu}^{(k)}, \boldsymbol{\sigma}^{2(k)} \}$. Inside $L(\eta;\eta^{(k-1)})$, since the denominator $P(O|\eta^{(k-1)})$ is a constant with respect to $\eta^{(k)}$,  the M step only needs to maximize 
\begin{flalign}
&L(\eta;\eta^{(k-1)})'= \nonumber \\
&\sum_{E^T \in \mathscr{L}^T}\sum_{C^T \in \mathscr{C}^T} P(E^T, C^T, O|\eta^{(k-1)})\log P(E^T, C^T, O|\eta)
\end{flalign}
over the parameter. However, based on the structure of Fig. \ref{Fig1}, the log-likelihood can be decomposed as
\begin{flalign}
&\quad \log P(E^T, C^T, U^T, Y^T| \eta)  \nonumber \\
&=\log \{ P(U^T, Y^T|E^T, C^T,\eta)P(E^T, C^T|\eta)\} \nonumber \\
&=\log\{P(U^T|E^T, \eta)P(Y^T|E^T, C^T, \eta)P(C^T|E^T,\eta)P(E^T|\eta)\} \nonumber  \\
&=\log P(U^T|E^T, \eta) + \log P(Y^T|E^T, C^T, \eta) \nonumber \\
&\qquad \qquad \qquad  + \log P(C^T|E^T,\eta) + \log P(E^T|\eta),
\end{flalign}
where each element depends on different optimization decision variables and there are no decision variables that have been shared by any two elements above. This implies that the maximization over the entire log-likelihood could be factorized into several independent subproblems, while the summation over all the optimal subproblem results leads to the optimal result of the entire problem. Mathematically, it indicates 

\begin{small}
\begin{flalign}
&\quad \max_{\eta} L(\eta;\eta^{(k-1)})' \nonumber \\
&=\max_{\bold{D}}\sum_{E^T \in \mathscr{L}^T}\sum_{C^T \in \mathscr{C}^T} P(E^T, C^T, O|\eta^{(k-1)})\log P(U^T|E^T, \eta) \nonumber \\
&+\max_{\pi^E,\bold{A}}\sum_{E^T \in \mathscr{L}^T}\sum_{C^T \in \mathscr{C}^T} P(E^T, C^T, O|\eta^{(k-1)})\log  P(E^T|\eta) \nonumber \\
&+\max_{\boldsymbol{\pi}^C, \bold{B}}\sum_{E^T \in \mathscr{L}^T}\sum_{C^T \in \mathscr{C}^T} P(E^T, C^T, O|\eta^{(k-1)})\log  P(C^T|E^T,\eta) \nonumber \\
&+\max_{\boldsymbol{\mu},\boldsymbol{\sigma}^2}\sum_{E^T \in \mathscr{L}^T}\sum_{C^T \in \mathscr{C}^T} P(E^T, C^T, O|\eta^{(k-1)})\log P(Y^T|E^T, C^T, \eta).
\end{flalign}
\end{small}
By solving the four subproblems above, the optimal solution for $\eta^{(k)}$ could be achieved, which is then used for the  next iteration if the convergence requirement is not yet satisfied.

\paragraph{Optimal $\bold{D}^{(k)}$} First, let us focus on learning the parameter $\bold{D}$, which is the emission probability of the PU energy arrival process. Since $\bold{D}=\{ d_{ij}\}$, learning $d_{ij}$ for all  $i,j \in \mathscr{L}$ is sufficient to have the estimated value of $\bold{D}$. Before moving on to the maximization procedure, the objective function could be further simplified as 
\begin{flalign}
\max_{\bold{D}} \sum_{E^T \in \mathscr{L}^T}P(E^T, U^T| \eta^{(k-1)})\log P(U^T|E^T,  \eta).
\end{flalign}
This is due to the fact that the hidden channel states $C^T$ could be marginalized by summing over all the possible outcome, and also
\begin{flalign*}
& \quad P(E^T,  U^T,Y^T| \eta^{(k-1)})\\ &=P(U^T,E^T|Y^T, \eta^{(k-1)})P(Y^T| \eta^{(k-1)}) \\
&=P(U^T|E^T, \eta^{(k-1)})P(E^T|Y^T,  \eta^{(k-1)})P(Y^T| \eta^{(k-1)}),
\end{flalign*}
where $P(Y^T| \eta^{(k-1)})$, $P(E^T|Y^T,  \eta^{(k-1)})$ are constant values with respect to $\bold{D}$. Besides,  $P(E^T| \eta^{(k-1)})$ is positive and independent of the decision variable $\bold{D}$, such that its multiplication with the objective function will not affect the estimation of $\bold{D}$. As such we consider joint probability $P(E^T, U^T| \eta^{(k-1)})$ instead of the conditional probability $P(U^T|E^T, \eta^{(k-1)})$ without influencing the maximization step. 

On the other hand, since $d_{ij}$ is the emission probability, the objective function needs to be decomposed from a joint probability into $T$  independent emission probabilities, in order to estimate the value of $d_{ij}$. The decomposition procedure is 

{\small\begin{flalign}
&\qquad \sum_{E^T \in \mathscr{L}^T}P(E^T, U^T| \eta^{(k-1)})\log P(U^T|E^T,  \eta) \nonumber \\
&=\sum_{E^T \in \mathscr{L}^T}P(E^T, U^T| \eta^{(k-1)})\log \prod_{t=1}^{T} P(U_t|E_t, \eta) \nonumber \\
&=\sum_{t=1}^T \sum_{i=0}^{L-1}\sum_{E^T \in \mathscr{L}^T}P(U^T,E^T| \eta^{(k-1)})\delta(E_t-i) \log P(U_t|E_t=i, \eta) \nonumber \\
&=\sum_{t=1}^T \sum_{i=0}^{L-1} P(U^T, E_t=i| \eta^{(k-1)}) \log P(U_t|E_t=i, \eta) \nonumber \\
&=\sum_{t=1}^T \sum_{i=0}^{L-1}  P(U^T, E_t=i| \eta^{(k-1)}) \log \sum_{j=0}^{L-1}\delta(U_t-j)d_{ij}.
\end{flalign}}
Inside the equation, $\delta(x)$ is the indicator function such that $\delta(x)=1$ when $x=0$ and $\delta(x)=0$ for any other values of $x$. Since $d_{ij}$ denotes the emission probability of the PU energy level, for any  $i$ the summation of $d_{ij}$ over all the possible outputs equals to one, i.e., $\sum\limits_{j=0}^{L-1}d_{ij}=1, \forall i$. Therefore, the maximization problem is finalized as 
\begin{flalign}
\max_{\bold{D}} &\sum\limits_{t=1}^T \sum\limits_{i=0}^{L-1}  P(U^T, E^t=i|\eta^{(k-1)}) \log \sum_{j=0}^{L-1}\delta(U_t-j)d_{ij} \nonumber \\
&\text{s.t.}  \sum_{j=0}^{L-1}d_{ij}=1
\end{flalign} 
This is a typical optimization problem to find a maximum of the objective function subject to an equality constraint, which could be solved by the Lagrange multiplier method \cite{boyd2004convex}.  After calculations, the optimal result for $d_{ij}^{(k)}$ is given as
\begin{flalign}
d_{ij}^{(k)}&=\frac{\sum\limits_{t=1}^T \delta(U_t-j)P(U^T, E_t=i|\eta^{(k-1)})}{\sum\limits_{t=1}^{T}P(U^T, E_t=i|\eta^{(k-1)})} \nonumber \\
&=\frac{\sum\limits_{t=1}^T \delta(U_t-j)\gamma_i^U(t)}{\sum\limits_{t=1}^T \gamma_i^U(t)} ,\quad i,j \in \mathscr{L}.
\end{flalign}
which implies that the estimated probability of emitting $j$ at state $i$ for the PU energy level equals the number of times that the output from the latent state $i$ is $j$ divided by the total number of times that the latent state is $i$.

\paragraph{Optimal $\pi^{E(k)}$, $\bold{A}^{(k)}$} The objective function for this subproblem could be further decomposed into two parts, in which the decision variables are $\pi^E$ and $\mathbf{A}=\{a_{i,j}\}$, respectively. The decomposed objective function is 
\begin{flalign}
 &\sum_{E^T \in \mathscr{L}^T}\sum_{C^T \in \mathscr{C}^T} P(E^T, C^T, U^T, Y^T|\eta^{(k-1)})\log P(E^T| \eta) \nonumber \\
&=\sum_{E^T \in \mathscr{L}^T}  P(E^T, U^T, Y^T| \eta^{(k-1)})\log P(E^T| \eta) \nonumber %\\
\end{flalign}
\begin{flalign}
&=\sum_{E^T \in \mathscr{L}^T}  P(E^T, U^T, Y^T| \eta^{(k-1)})  \nonumber \\
&\qquad \qquad \qquad \qquad \times \left\{\log \pi^{E} +\sum_{t=2}^T  \log a_{E_{t-1} E_t}\right\}.
\end{flalign}
 Clearly, the optimization procedures to maximize the objective function over $\pi^E$ and $\bold{A}$ could be treated as two independent processes. Therefore, after factorizing the joint probablities into $T$  independent probabilities based on the Markovian property, the problems become
\begin{flalign}\label{sub_1}
\max_{\pi^{E}_{i}}& \sum_{i\in \mathscr{L}}  P(U^T, Y^T, E_1=i| \eta^{(k-1)}) \log \pi^{E}_{i} \nonumber \\
&\text{s.t.} \sum_{i=0}^{L-1}\pi^{E}_{i}=1, \quad i \in \mathscr{L},
\end{flalign}
and 
\begin{flalign}\label{sub_2}
\max_{a_{ij}}& \sum_{t=2}^T \sum_{i\in \mathscr{L}} \sum_{j\in \mathscr{L}} P(U^T, Y^T, E_{t-1}=i, E_t=j|\eta^{(k-1)}) \log a_{ij}     \nonumber \\
& \text{s.t.} \sum_{j=0}^{L-1} a_{ij}=1, \quad i \in \mathscr{L}.
\end{flalign}
By solving the subproblems with the Lagrange multipler method, the estimated values of $\pi^E$ and $\bold{A}$ for  step $k$ are 
\begin{flalign}
\pi_{i}^{E(k)}&=\frac{P(U^T, Y^T, E_1=i|\eta^{(k-1)})}{P(U^T, Y^T|\eta^{(k-1)})} \nonumber \\
&=\frac{\sum\limits_{j \in \mathscr{C}}{\gamma_{ji}^{U,Y}(1)}}{\sum\limits_{j \in \mathscr{C}}\sum\limits_{i \in \mathscr{L}} \gamma_{ij}^{U,Y}(1)},
\end{flalign}
and
\begin{flalign}
a_{ij}^{(k)}&=\frac{\sum\limits_{t=2}^TP(U^T, Y^T, E_{t-1}=i, E_{t}=j|\eta^{(k-1)})}{\sum\limits_{t=2}^T P(U^T, Y^T, E_{t-1}=i|\eta^{(k-1)})} \nonumber \\
&=\frac{\sum\limits_{t=2}^T \sum\limits_{m,l\in \mathscr{C}} \varepsilon_{mi,lj}^{U,Y}(t-1)}{\sum\limits_{t=2}^{T}\sum\limits_{m\in\mathscr{C}}\gamma_{mi}^{U,Y}(t-1)},
\end{flalign}
for any  $i \in \mathscr{L}$. % If the performance can be sacrificed  to simplify the algorithm, estimating the transition probability for $\{E_t\}$ could neglect the influence of $\{ Y_t\}$, i.e., only considering the joint probability  $P(U^T, E^T|\eta^{(k-1)})$ in the optimization procedure. Then the estimation for $\{E_t\}$ can directly use the result of Baum-Welch algorithm, since the PU energy level Markov chain becomes the traditional HMM. 
It can be seen that as channel and energy observations contain the information of the PU energy level, they both contribute to the estimation of the PU energy transition matrix. 

\paragraph{Optimal $\boldsymbol{\pi}^{C(k)}, \bold{B}^{(k)}$} Following the previous arguments, the objective function here can also be decomposed as 
{\small\begin{flalign}
&\quad \sum_{E^T \in \mathscr{L}^T}\sum_{C^T \in \mathscr{C}^T} P(U^T, Y^T, C^T, E^T|\eta^{(k-1)}) \log P(C^T| E^T, \eta) \nonumber \\
&=\sum_{E^T \in \mathscr{L}^T}\sum_{C^T \in \mathscr{C}^T}  P(U^T|E^T,\eta^{(k-1)})P(E^T, C^T, Y^T|\eta^{(k-1)}) \nonumber \\
&\qquad \qquad \qquad \quad \times \left\{\log \pi^C+ \sum_{t=1}^{T} \log b_{C_{t-1}E_t, C_t}\right\}.
\end{flalign}}
Since $P(U^T|E^T,\eta^{(k-1)})$ is not changing with respect to $\boldsymbol{\pi}^C$ and $\bold{B}$, it can be neglected for estimating these parameters. After factorizing the objective function into $T$ independent elements, the optimal parameters are obtained as 
{\begin{flalign}
\pi_{{ij}}^{C(k)}&=\frac{P(Y^T, E_1=i,C_1=j|\eta^{(k-1)})}{P(Y^T, E_1=i|\eta^{(k-1)})} \nonumber \\
&=\frac{\gamma_{ji}^Y(1)}{\sum\limits_{j\in \mathscr{C}} \gamma_{ji}^Y(1)}, \quad i\in \mathscr{L}, j \in \mathscr{C},
\end{flalign}}
and 
\begin{flalign}
b_{ij,k}^{(k)}&=\frac{\sum\limits_{t=2}^{T}P(Y^T, C_{t-1}=i, E_t=k, C_t=j|\eta^{(k-1)})}{\sum\limits_{t=2}^{T}P(Y^T, C_{t-1}=i, E_t=k|\eta^{(k-1)})} \nonumber \\
&=\frac{\sum\limits_{t=2}^{T}\sum\limits_{m=0}^{L-1}\varepsilon_{im,jk}^Y(t-1)}{\sum\limits_{t=2}^{T} \sum\limits_{m=0}^{L-1}\sum\limits_{j\in \mathscr{C}}\varepsilon^Y_{im,jk}(t-1)}, \quad i,j \in \mathscr{C}, k \in \mathscr{L}.
\end{flalign}

\paragraph{Optimal $\boldsymbol{\mu}$, $\boldsymbol{\sigma}^2$} The channel output $\{ Y_t\}$ is a continous states process, such that given the latent states of channel and PU energy, $\{ Y_t\}$ follows a Gaussian distribution for any $t$. Learning  $\boldsymbol{\mu}=\{ \mu_{{ij}}\}$ and variance $\boldsymbol{\sigma}^2=\{\sigma^2_{{ij}}\}$ for $i \in \mathscr{C}$ and $j \in \mathscr{L}$ of the conditional Gaussian distribution could determine the emission probabilities of channel and PU energy states.  

Conditioned on the hidden PU energy process, the observation of the SU energy level is independent of the hidden channel state and the channel output; therefore, $U^T$ does not contribute to the estimation of the channel output parameters. Since given the latent states, the channel output $\{ Y_t\}$ is independent over time, we have
{\small\begin{flalign}
&\quad \sum_{E^T}\sum_{C^T} P(Y^T, C^T, E^T|\eta^{(k-1)})\log P(Y^T|E^T,C^T,\eta) \nonumber \\
&=\sum_{E^T}\sum_{C^T}  P( Y^T, C^T, E^T|\eta^{(k-1)})\log \prod_{t=1}^T P(Y_t|E_t,C_t,\eta) \nonumber \\
&=\sum_{t=1}^T \sum_{i \in \mathscr{C}} \sum_{j\in \mathscr{L}}P( Y^T, C_t=i, E_T=j|\eta^{(k-1)}) \nonumber  \\
& \quad \qquad \qquad \qquad \qquad \times \log P(Y_t|C_t=i, E_t=j, \eta),
\end{flalign}}
where $P(Y_t|C_t=i, E_t=j, \eta) \sim N(Y_t; \mu_{{ij}}, \sigma^2_{{ij}})$ and $\sum\limits_{E^T},\sum\limits_{C^T}$ stand for $\sum\limits_{E^T \in \mathscr{L}^T}$,$\sum\limits_{C^T \in \mathscr{C}^T}$ respectively. Since this is an unconstrained convex optimization problem with a differentiable objective function,  letting the first derivative of the objective function equals to zero and solving the rest will provide the optimal point for $\mu_{{ij}}$ and $\sigma^2_{{ij}}$. Then the estimated values of these parameters are 
{\begin{flalign}
\mu_{{ij}}^{(k)}&=\frac{\sum\limits_{t=1}^TP(Y^T, C_t=i, E_t=j|\eta^{(k-1)})Y_t}{\sum\limits_{t=1}^T P(Y^T, C_t=i, E_t=j| \eta^{(k-1)})}
=\frac{\sum\limits_{t=1}^T \gamma_{ij}^Y(t)Y_t}{\sum\limits_{t=1}^T \gamma_{ij}^Y(t)}
  %\\
\end{flalign}}
%\begin{flalign}
%&=\frac{\sum\limits_{t=1}^T \gamma_{ij}^C(t)Y_t}{\sum\limits_{t=1}^T \gamma_{ij}^C(t)}
%\end{flalign}
and 
{\begin{flalign}
\sigma_{ij}^{2(k)}&=\frac{\sum\limits_{t=1}^TP(Y^T, C_t=i, E_t=j|\eta^{(k-1)})(Y_t-\mu_{{ij}}^{(k)})^2}{\sum\limits_{t=1}^T P(Y^T, C_t=i, E_t=j| \eta^{(k-1)})} \nonumber \\
&=\frac{\sum\limits_{t=1}^T \gamma_{ij}^Y(t)(Y_t-\mu_{{ij}}^{(k)})^2}{\sum\limits_{t=1}^T \gamma_{ij}^Y(t)}.
\end{flalign}}
However, the expressions above only work for $i =1$, $\forall j\in \mathscr{L}$. Since when $i =0$, PU does not transmit signal and the channel output does not contain any information about the PU energy level. As we mentioned, in this case  $\mu_{{ij}}$ and variance $\sigma^2_{{ij}}$ are identical for any $ j \in \mathscr{L}$. %And the estimated parameters are 
%$\mu_{c_{ij}}$ and $\sigma^2_{c_{ij}}$. 
Then the estimated values of these parameters are 
\begin{flalign}
\mu_{{ij}}^{(k)}&=\frac{\sum\limits_{t=1}^TP(Y^T, C_t=i|\eta^{(k-1)})Y_t}{\sum\limits_{t=1}^T P(Y^T, C_t=i| \eta^{(k-1)})} \nonumber \\
&=\frac{\sum\limits_{t=1}^T \left\{ \sum\limits_{j=0}^{L-1} \gamma_{ij}^Y(t) \right\} Y_t}{\sum\limits_{t=1}^T \sum\limits_{j=0}^{L-1}\gamma_{ij}^Y(t)}
\end{flalign}
and 
\begin{flalign}
\sigma_{ij}^{2(k)}&=\frac{\sum\limits_{t=1}^TP(Y^T, C_t=i|\eta^{(k-1)})(Y_t-\mu_{{ij}}^{(k)})^2}{\sum\limits_{t=1}^T P(Y^T, C_t=i| \eta^{(k-1)})} \nonumber \\
&=\frac{\sum\limits_{t=1}^T \left\{  \sum\limits_{j=0}^{L-1} \gamma_{ij}^Y(t)\right\}(Y_t-\mu_{{ij}}^{(k)})^2}{\sum\limits_{t=1}^T  \sum\limits_{j=0}^{L-1}\gamma_{ij}^Y(t)},
\end{flalign}
when $i =0$.

Based on the calculations above, the estimated parameter could be updated to $\eta^{(k)}$. As the algorithm is derived from EM algorithm, the expected log-likelihood is non-decreasing over iterations. Let $\epsilon$ be the acceptable difference for convergence, if $L(\eta;\eta^{(k)})-L(\eta;\eta^{(k-1)})\leq \epsilon$, we claim $\eta^{(k)}$ as the final estimated value for parameter $\eta$ and use it for further applications; otherwise, we use $\eta^{(k)}$ to calculate the expectation of the log-likelihood and continue EM iteration to get $\eta^{(k+1)}$.

%To apply original Baum-Welch algorithm for parameter estimation of this model, we can vectorize the hidden state and observation, i.e., the PU energy state and channel state will be treated as a vectorized hidden state with $L \times M$ numbers of different states and the observations of SU energy state and channel output are 

\subsection{Initialization}
As can be seen from \cite{543975}, since at every iteration the estimated parameter increases the expected log-likelihood function value, the proposed algorithm based on the EM algorithm converges to a stationary point, which can be a local optimum or a saddle point. In other words, the global maximum is not guaranteed such that convergence to which local optimum depends on the initialization of the algorithm. 

%There are some existing methods that provide a better (in some sense) initializations for the EM algorithm. A subtle method is introduced in \cite{hsu2012spectral}, where a spectral condition is required on the observation and transition matrices. This work performs singular value decompositions over the first, second and third order of observation probabilities to learn the parameters; but it does not recover the transition and observation models directly. In addition to above computation burden, it also requires certain transformations to obtain the initial values of the proposed algorithm. Moreover, this method requires several short sequences of observation rather than one long sequence of observation, which makes it inappropriate to the model we have.

For simplicity, a multi-try random initialization routine is adopted in our algorithm. Since at each time the algorithm with a random initialization converges to a local optimum, it is logical to run the algorithm multiple times with different initializations and choose the outcome that has the maximum likelihood value. Thorough discussions of optimal initialization strategies is out of the scope of this paper.

In summary, the proposed algorithm for the HIMM parameter estimation is illustrated below. 
\begin{codebox}
\Procname{$\proc{Parameter Estimation Algorithm}$}
\li Define iteration variable $k=1$. Take the initialization\\ value $\eta^{(0)}= \{\pi^{E(0)}, \boldsymbol{\pi}^{C(0)}, \bold{A}^{(0)}, \bold{B}^{(0)}, \bold{D}^{(0)}, \boldsymbol{\mu}^{(0)}, \boldsymbol{\sigma}^{2(0)} \}$ \\ from certain distributions.
\li    Use $\eta^{(k-1)}$ to calculate intermediate variables \\ $\alpha^U, \beta^U, \gamma^U, \varepsilon^U,\alpha^Y, \beta^Y, \gamma^Y, \varepsilon^Y, \alpha^{U,Y}, \beta^{U,Y}, \gamma^{U,Y}, \varepsilon^{U,Y}$.
\li      Use the intermediate variables from step $2$ to calculate\\ the current estimated parameter $\eta^{(k)}$.
\li    Based on $\eta^{(k)}$, calculate the expected log-likelihood\\ $L(\eta; \eta^{(k)})$.
\li      Repeat steps $2$ to $4$ until $L(\eta;\eta^{(k)})-L(\eta;\eta^{(k-1)})\leq \epsilon$.
\li      Repeat steps $1$ to $5$ with different initializations to find\\ $\eta^{(k)}$ with the largest likelihood value.
\li       The estimated parameter is set ${\eta}'=\eta^{(k)}$.
\end{codebox}

\section{Simulation Results}
In this section, numerical results are provided to validate the effectiveness of the proposed parameter learning and 2-D sensing algorithms. Our scheme considers both channel and energy observations while the reference energy detector based spectrum sensor, which is one of the well-accepted traditional PU detection methods, only uses channel observations. 

\begin{figure}[!t]
\centering
\includegraphics[width=3.5in]{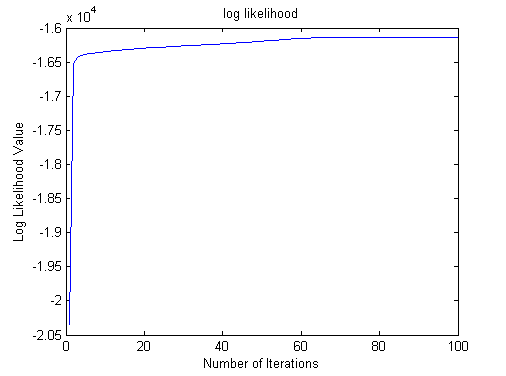}
\caption{Increasing log-likelihood with Parameter Learning}
\label{log-likelihood}
\end{figure}

\begin{figure}[!t]
\centering
\includegraphics[width=3.5in]{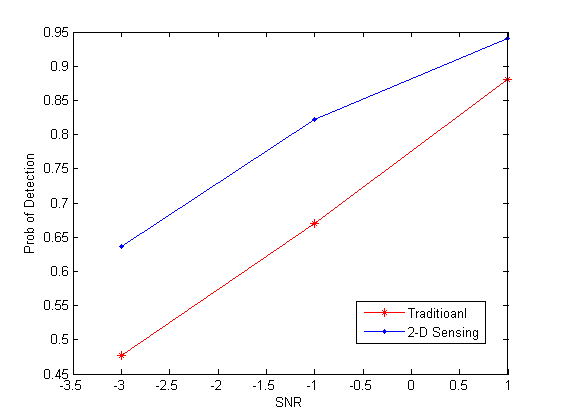}
\caption{Comparison of  Detection Performance}
\label{afterlearning}
\end{figure}

For the parameter learning, we consider a training model with $5000$ channel outputs and SU energy observations. The trend of the increasing log-likelihood with the proposed parameter learning algorithm is showed in Fig. \ref{log-likelihood}. It can be seen that there is a sharp log-likelihood increase in the first five iterations, and after $60$ steps the algorithm converges. As one property of the EM algorithm, the log-likelihood converges monotonically and every random initial value guarantees the convergence. In this simulation, we tried $15$ different initial values, and the estimated parameters are obtained by choosing the one with the maximum converged log-likelihood value.  The initial values are randomly chosen according to the uniform distribution between $0$ and $1$ with proper normalizations.

In Fig. \ref{afterlearning}, we show the differences of probability of detection versus SNR at the SU between the traditional sensing and the proposed 2-D sensing methods, with HIMM parameters learned by the proposed learning algorithm. The SNR at the SU is defined as the SU received signal power (from PU) divided by the channel noise power.  In this figure, the channel noise level is set as $\sigma_n^2=3$ dBw, and the channel path loss is $-4$ dB. The energy state takes values from $\{1,2,3,4 \}$, in which the insufficient energy state subset is $\mathscr{L}_0=\{ 1 \}$. 
Note that with each SNR value, we set the two schemes in comparison to have the same false alarm performance.

From Fig. \ref{afterlearning}, we see that the 2-D sensing method outperforms the traditional energy detector based scheme over all SNR values, since the 2-D method considers observations of both the channel state and the PU energy state, and it also utilizes the hidden Markov signaling structure between the PU and the SU. As the estimator used to estimate the hidden channel and energy states is a MAP estimator, it actually provides the optimal result that jointly estimates the hidden states of the Markov model.  When the SNR is low, the observation over the PU-to-SU channel is not reliable for PU detection, which largely affects the performance of the traditional  method. Since 2-D sensing also takes the energy observation into account, which is not affected by the PU-to-SU channel,  the performance of 2-D sensing is much better. As SNR goes up, the channel observations become more reliable and the benefit of using the additional energy information is less, which implies that the relative gain of using 2-D sensing will decrease. However, as the 2-D sensing method does not sacrifice any information during the estimation, although the gain may become less, 2-D sensing still outperforms the traditional method in the high SNR region.

\begin{figure}[!t]
\centering
\includegraphics[width=3.5in]{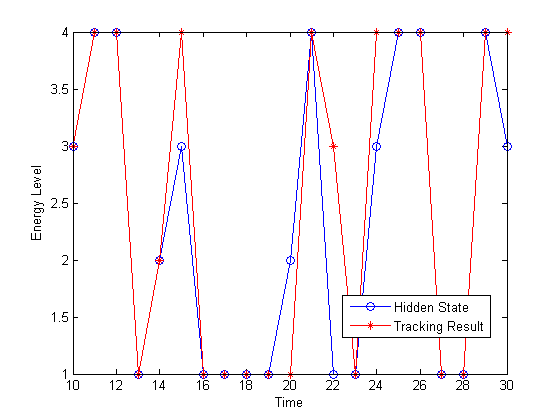}
\caption{Tracking Performance of Energy States}
\label{Tracking Result}
\end{figure}

Another advantage of using the proposed 2-D sensing method is that SU could estimate the transmit power of PU, which is proportional to $E_t$ for broader CR applications \cite{5054703}\cite{6477928}. In Fig. \ref{Tracking Result}, we show the tracking performance of the hidden PU energy state $E_t$, where it can be seen that the proposed method could estimate the hidden energy state quite well.

\section{Conclusion}
In this paper, we applied a HIMM model to represent the interaction between energy harvesting based primary and cognitive radios, and provided an EM based algorithm to estimate the parameters in this HIMM structure. Then we proposed a novel 2-D sensing scheme, which jointly considers the observations from the spectrum and power dimensions. The proposed scheme could sense the spectrum and estimate the energy level for PU transmission simultaneously.  We showed that the proposed 2-D sensing method outperforms the traditional spectrum sensing method, since it utilizes the facts that the PU power usage and channel usage are interdependent events, and the PU/SU energy harvesting processes are spatially correlated.

\bibliographystyle{IEEEtran}
\bibliography{IEEEabrv,reference}
%\end{spacing}
\end{document}